\documentclass[%
 reprint,
 amsmath,amssymb,
 aps,
]{revtex4-2}

\usepackage{graphicx}
\usepackage{dcolumn}
\usepackage{bm}

\begin{document}

\preprint{APS/123-QED}

\title[]{Epsilon-near-zero nanoparticles}
\author{Ibrahim Issah}
\author{Jesse Pietila}
\author{Tommi Kujala}
\affiliation{%
 Tampere University, Faculty of Engineering and Natural Sciences, 33720 Tampere, Finland\\
}%
\author{Matias Koivurova}
\affiliation{
  Tampere University, Tampere Institute for Advanced Study, 33100 Tampere, Finland\\
}%
\affiliation{
 Tampere University, Faculty of Engineering and Natural Sciences, 33720 Tampere, Finland\\
}%

\author{Humeyra Caglayan}%
\email {humeyra.caglayan@tuni.fi}

\author{Marco Ornigotti}%
\email { marco.ornigotti@tuni.fi}

\affiliation{ 
Tampere University, Faculty of Engineering and Natural Sciences, 33720 Tampere, Finland}%

\date{\today}

\begin{abstract}
In this work, we propose epsilon-near-zero (ENZ) nanoparticles formed of metal and dielectric bilayers and employ the effective medium approach for multilayered nanospheres to study their optical response. 
We obtained a passive tunable ENZ region by varying the radii of the proposed bilayer nanospheres, ranging from visible to near-IR. In addition, we present the absorption and scattering cross-section of ENZ nanoparticles using an open-source, transfer-matrix-based software (STRATIFY). 
The proposed ENZ nanoparticle is envisioned to be experimentally realized using chemical synthesis techniques. 
\end{abstract}

\maketitle

\section{\label{sec:level1}Introduction}
Epsilon-near-zero (ENZ) materials exhibit a dielectric permittivity approaching zero at a frequency close to the material's plasma frequency \cite{Alu2007, Zheng2021}. Transparent conductive oxides such as indium tin oxide (ITO) are naturally occurring ENZ materials with ENZ wavelengths in near-infrared and mid-infrared regions \cite{Naik2011}. A metamaterial composed of alternating layers of metal and dielectric was also demonstrated to exhibit ENZ properties in the visible region \cite{gao2013experimental, maas2013experimental}.


Notably, ENZ materials studied by many authors consist of nanostructures or meta-atoms, which require fabrication techniques such as focused ion beam (FIB) milling, laser ablation, atomic layer deposition (ALD), and electron-beam (e-beam) lithography \cite{https://doi.org/10.1002/adom.202200081, Vesseur2013}.
On the other hand, the synthesis of nanoparticles (NPs) and their incorporation into materials are among the most studied topics in chemistry, physics, and material science. Furthermore, it has been demonstrated that localized surface plasmon (LSPR) depends on the size, shape, and material of the nanoparticles. For example, small Au NPs (between 5 and 10 nm) have the LSPR band around 520 nm, while for bigger particles (between 50 and 100 nm), this peak is red-shifted up to 570 nm. Other materials such as TCOs, transition metal nitrides \cite{GULER2015227}, organic conductive materials \cite{Yang2019}, and highly doped semiconductors have also been identified to exhibit plasmonic behavior at different spectral regions. 

In addition, nanoparticles composed of a metal-dielectric complex have shown to exhibit interesting light-matter interactions with  many vital applications in physics and chemistry, such as scattering \cite{Graf2002, Hasegawa2006} and nonlinear optics \cite{Scherbak2018, Pu2010}, sensing \cite{Ochsenkuhn2009,Jain2007}, fluorescence \cite{Zhang2019, Chan1991}, and up-conversion enhancement \cite{Zhang2010}, surface plasmon amplification \cite{Zuev2010, Noginov2009}, hydrogen generation \cite{Curtis2021}, and solar energy harvesting \cite{Rativa2018, Phan2018}.

Although planar films made of similar materials have been investigated as ENZ medium for many applications, nanoparticles have not been considered ENZ materials. 
We envision that the ENZ nanoparticles could provide a new design solution for low-cost tunable ENZ materials, which have wide prospects for application in photonics.

In this work, we propose the possibility of the utilization of nanoparticles as ENZ materials. To verify this approach, we employ the effective medium approach to model the optical response of a multilayer sphere as an effective bulk spherical medium. In particular, by varying the structural properties of the nanospheres, we show how the ENZ character of such multilayered nanoparticles can be easily tuned from the visible (VIS) to the near-IR (NIR) region of the electromagnetic spectrum. 

\begin{figure}
\includegraphics[width=1\linewidth]{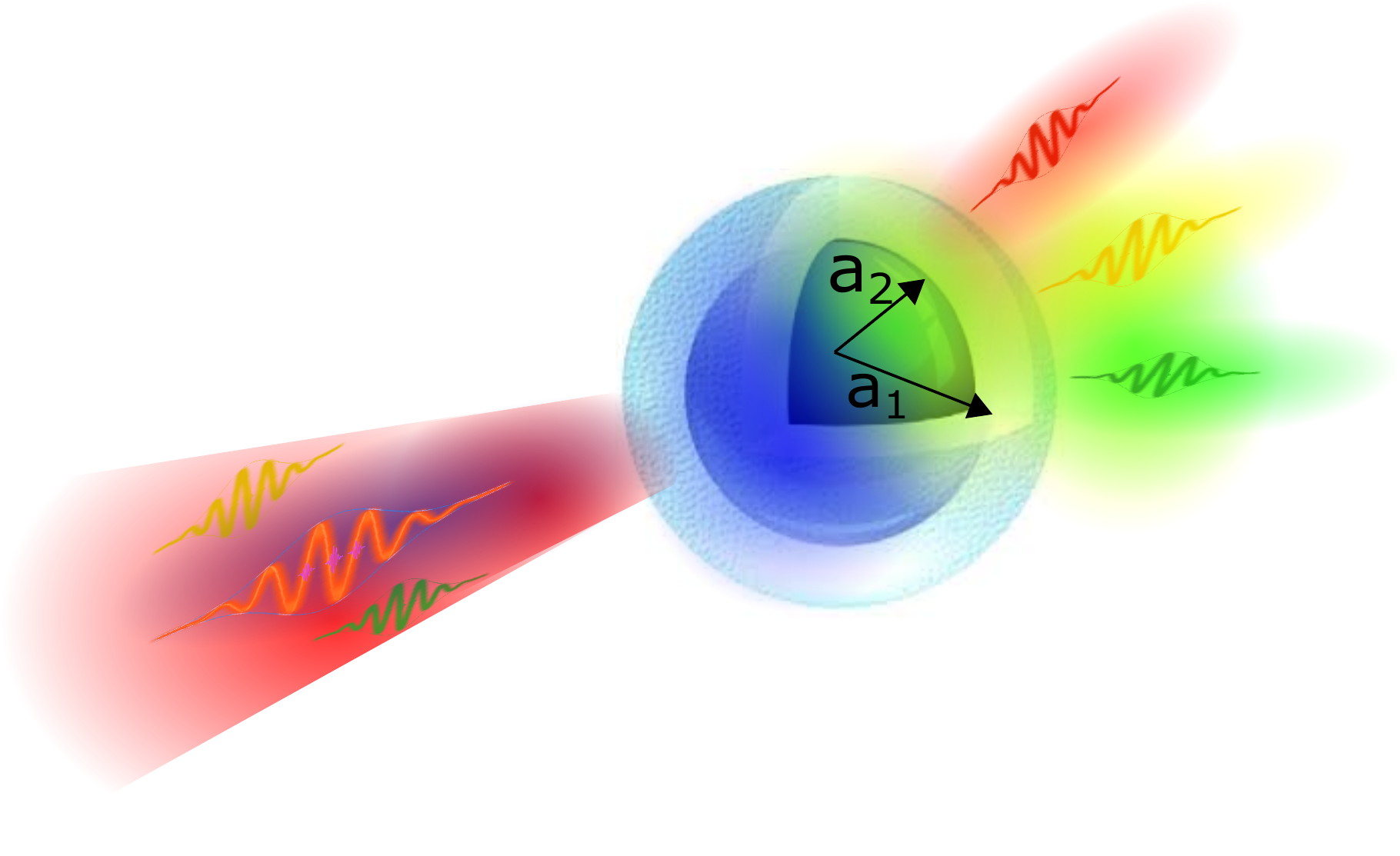}
\caption{Schematic representation of the bi-layer spherical nanoparticles. $a_{2}$ and $a_{1}$ represent the inner and outer radius of the sphere. The inner, dark blue, sphere represents the dielectric core (SiO\textsubscript{2}) of the nanoparticle, while the outer, light blue, shell represents the silver coating.} 
\label{fig1}
\end{figure}

\section{\label{sub:Design}Structure design and Modeling}
The ENZ nanoparticles we considered in this work, are bi-layer structures consisting of an inner dielectric core, for which we employ silicon dioxide (SiO$_2$), and an outer metallic shell, as shown schematically in Fig. \ref{fig1}. Due to the high carrier concentration, electron mobility, and strong electromagnetic field confinement property of noble metals \cite{Blaber2010}, we choose silver (Ag) for the outer shell.

To analyze the absorption and scattering of the aforementioned ENZ nanoparticles, we use a transfer-matrix approach (based on the open-source STRATIFY code \cite{Rasskazov2020}). 
%
%
As a sanity check, we also implemented the optical response (i.e., scattering and absorption) for different materials 
and identified that the Ag-SiO$_2$ bilayer structure gives us the required ENZ region of interest (i.e., visible to NIR). First, we focused on the effective ENZ optical responses of a bi-layer spherical nanoparticle embedded within a host medium (i.e., air), and later extended the effective permittivity approach to multilayer nanosphere composites.

The wavelength-dependent complex dielectric functions for Ag and SiO$_2$ were taken from the material data of Johnson and Christy \cite{Johnson1972}. The scattering, absorption, and electromagnetic near-field distribution are calculated using STRATIFY (i.e., recursive transfer matrix method (RTMM) MATLAB code). 


\section{\label{sec:level2}Theory}
In this section, we briefly present the theoretical background needed to derive the effective permittivity of the bi-layer spherical nanoparticle. Our approach makes use of an effective description of the electric permittivity of a metal-dielectric layered nanoparticle, following the methods presented in Ref. \citenum{Chen1998}. In particular, we assume that a bilayer (and, by simple generalization, a multilayer) nanoparticle can be seen as a composite material, where the outer layer (the shell) plays the role of the host medium, whereas the inner layer (the core) is the inclusion. This assumption allows us to use the theory of effective media to simplify the complex problem of a multi-layered spherical nanoparticle, reducing it to the simpler one of a single, homogeneous spherical nanoparticle, described via an effective permittivity. 

For the case of a metal-dielectric nanoparticle as the one depicted in Fig. \ref{fig1}, for example, this can be easily done using standard methods of electrostatics, ie., by placing the nanoparticle in a homogeneous electric field and solving the Laplace equation for the electrostatic potential in the whole space \cite{Jackson:100964}. By using this procedure, one can easily show that the electric field generated by the bilayered sphere in the host medium (air, in the case of Fig. \ref{fig1}) is the same as that of a single sphere with radius $a_1$ and permittivity
\begin{equation}
\tilde{\epsilon_{1}} = \frac{1 - 2G}{1 + G}\epsilon_{1},
\label{eqn2}
 \end{equation}
 with
\begin{equation}
G = \frac{\epsilon_{1} - \epsilon_{2}}{2\epsilon_{1}+\epsilon_{2}}\left(\frac{a_{2}}{a_{1}}\right)^3,
\label{eqn3}
 \end{equation}
with $\epsilon_1$ being the permittivity of the shell, $\epsilon_2$ the permittivity of the core, and $a_{2}$ the radius of the inner sphere.
 

The above theory can then be generalized for multilayered structures by expressing the multilayered permittivities by simply applying Eq. \eqref{eqn2} recursively to each couple of layers, from the outer to the inner one \cite{Chen1998}, thus obtaining, for the general $k$-th layer 
  \begin{equation}
     \tilde{\epsilon_{k}} = \frac{1 - 2G_{k}}{1 + G_{k}}\epsilon_{k},
     \label{eqn4}
 \end{equation}
 with
 \begin{equation}
     G_{k} = \frac{\epsilon_{k} - \tilde{\epsilon}_{k+1}}{2\epsilon_{k}+\tilde{\epsilon}_{k+1}}\left(\frac{a_{k+1}}{a_{k}}\right)^3.
 \end{equation}
 Here, the layers are numbered from outside to inside so that $k = 1$ represents the outer layer and $k = N$ represents the inner one. The permittivity for each layer $k = \{N - 1, N - 2, ..., 2, 1\}$ is expressed using the above equations. 

\section{\label{sec:level3}Results and Discussion}

  \begin{figure*}
  \centering
    \includegraphics[width=1\linewidth]{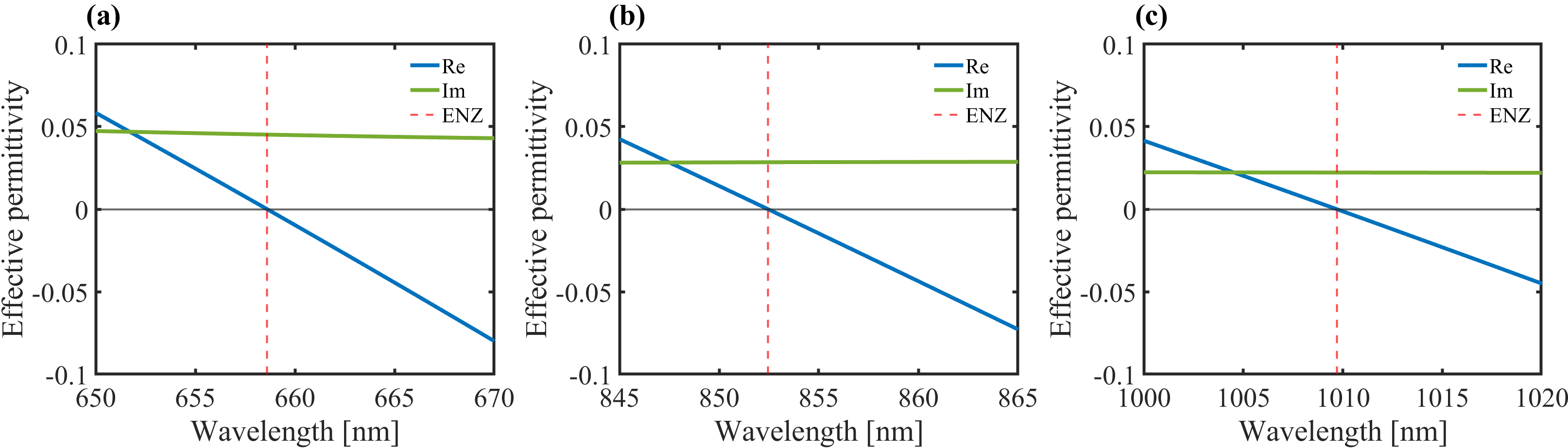}
    \caption{Effective permittivities of different bilayer spheres near their characteristic ENZ wavelengths, (a) $a_{2,1}=\{38,40\}$, (b) $a_{2,1}=\{68,70\}$, and (c) $a_{2,1}=\{98,100\}$ nm. (Re) and (Im) represents the real and imaginary part of the effective permittivity. Their corresponding ENZ regions are marked with red dashed lines.} 
    \label{fig2}
\end{figure*}

 \begin{figure*}
\includegraphics[width=1\linewidth]{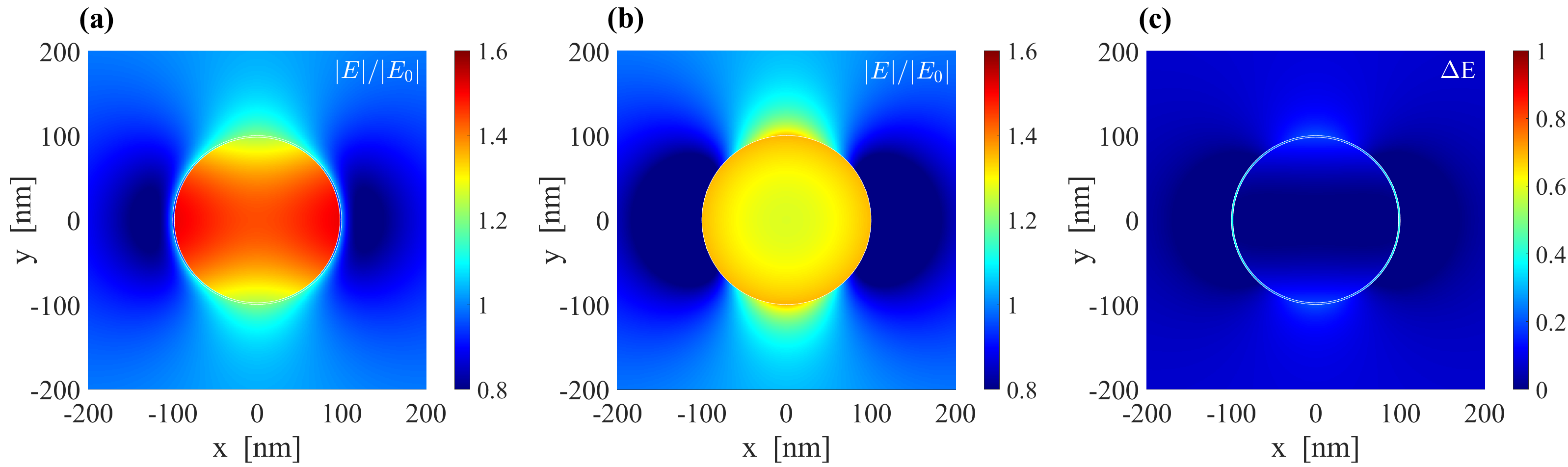}
\caption{The electric field for the bilayer structure, $a_{2,1} =\{98,100\}$ nm at the ENZ wavelength. (a) The actual bilayer structure and (b) the bulk effective medium structure. (c) The difference of the normalized near-field distribution between (a) and (b).}
\label{fig3}
\end{figure*}

From the above theory, we determined the effective permittivity for the proposed spherical bilayer structure to obtain the ENZ points as a function of the varied core diameters with outer layer thicknesses fixed at 2 nm. Figure \ref{fig2} shows the effective permittivities for the different ENZ nanoparticles with different core diameters (i.e., $a_{2,1}=\{38,40\}$, $a_{2,1}=\{68,70\}$, and $a_{2,1}=\{98,100\}$ nm, respectively). $a_{2,1}$ represents the radii of the inner and outer shells, respectively. We obtained different ENZ points by changing the radii of the bilayer nanospheres. 

In particular, for $a_{2,1}=\{38,40\}$ nm [Fig. \ref{fig2}(a)], the ENZ wavelength is found to be $\lambda_{ENZ}\approx$ 659 nm. Similarly, by changing the inner and outer radii of the ENZ nanoparticle, we observe a redshift of the ENZ wavelength to $\lambda_{ENZ}\approx$ 852 nm [Fig. \ref{fig2}(b)] and $\lambda_{ENZ}\approx$ 1010 nm [Fig. \ref{fig2}(c)], respectively. This is the first main result of our work which extends to the numerical calculations of the nanosphere's near-field enhancement, absorption and scattering cross-sections, as well as, exploiting the possibility of using the effective medium formulations of bilayer nanospheres into multilayer ENZ nanoparticles.

Ag-SiO$_2$ nanoparticles clearly show ENZ behavior in the VIS and NIR spectral regime. The position of the ENZ wavelength of such nanoparticles can be easily controlled by suitably tuning the inner core of the nanoparticle. In our simulations, it was found that the outer radius of the nanoparticle must obey the rule $a_1=<\lambda_{ENZ}/10$ for the effective medium approach to remain valid. 
In addition, it was identified that one should consider the parametric variations of the inner core and outer shell of the nanoparticles, as huge thickness variations between the two layers could affect the bilayer nanoparticle from exhibiting ENZ properties. This is due to the bilayer nanoparticle exhibiting properties of the outer shell instead of the complex media. As such, it is relevant to choose the right fill fraction of the nanoparticle to attain the required ENZ properties.
%


To verify the reliability of the effective medium approach described above, we compare the electromagnetic field distribution of the proposed effective medium theory and the bilayer structure. As it can be seen from Fig. \ref{fig3}, the electric field distribution outside the nanoparticle, calculated using STRATIFY, for both the case of a bi-layer structure [Fig. \ref{fig3}(a)] and a bulk sphere with effective permittivity $\epsilon_{eff}$ [Fig. \ref{fig3}(b)] at the ENZ wavelength, give relatively similar results, as can be seen from Fig. \ref{fig3}(c). 
%
Moreover, our numerical simulation increase in performance when using the effective medium approximation instead of the full bilayer structure of the nanoparticle. We also identified that a bilayer metal-dielectric structure is sufficient to obtain the desired ENZ properties. As a result, we focused on the particle with radius $a_{2,1} =\{98,100\}$ nm which can produce Rayleigh scattering since the particle size is smaller than the wavelength of the impinging electromagnetic field. 
%
Although a full numerical optimization could be performed, we only considered the ENZ nanoparticle with radius $a_{2,1} =\{98,100\}$ nm, as it shows the capabilities of our approach shown in Fig. \ref{fig3} and corresponds to experimentally plausible nanoparticles \cite{C4TC02780A}.

 

\begin{figure}[!ht]
\includegraphics[width=1\linewidth]{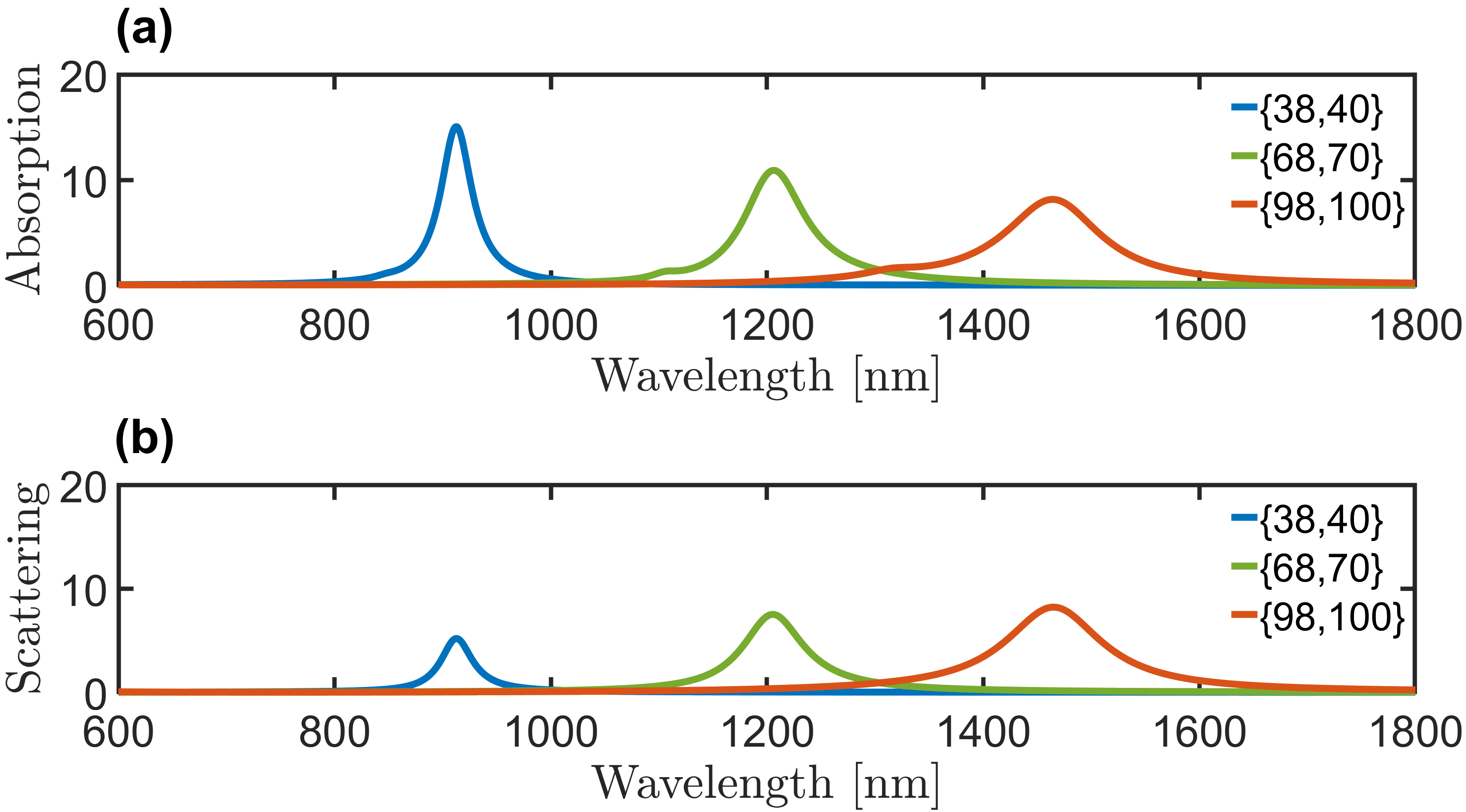}
\caption{(a) Absorption and (b) Scattering cross-sections for the bilayered spherical nanoparticles with radii $a_{2,1} =\{38,40\}$ nm,  $a_{2,1} =\{68,70\}$ nm, and $a_{2,1} =\{98,100\}$ nm. The values in the legend represent the radius values of the inner and outer spherical layers.} 
\label{fig4}
 \end{figure}

In addition, we numerically calculated the fundamental extinction cross-sections and the ENZ-dependent parametric sweep for different layers of the ENZ nanoparticle in the case of a plane wave incident on it using the effective medium formulation.  
Figures \ref{fig4} (a) and (b) show the absorption and scattering cross-sections for the three different ENZ nanoparticles, i.e.,  $a_{2,1} =\{38,40\}$ nm (blue),  $a_{2,1} =\{68,70\}$ nm (green), and $a_{2,1} =\{98,100\}$ nm (red), with absorption and scattering peak resonances at $\lambda\approx$ 913 nm, 1206 nm, and 1466 nm, respectively. 
The maximum extinction resonance wavelengths for the proposed ENZ nanoparticles are redshifted, compared to their characteristic ENZ wavelengths which are presented in Table \ref{table:1}. 

\begin{table}[!ht]
\centering
\begin{tabular}{|l|l|c|}
\hline
Nanoparticle size (nm) & $\lambda_{ENZ}$ & $\lambda_{Resonance}$\\ \hline
\{38, 40\}             & 659 nm       & 913 nm                                       \\ \hline
\{68, 70\}            & 852 nm       & 1206 nm                                       \\ \hline
\{98, 100\}           & 1010 nm      & 1466 nm                                      \\ \hline
\end{tabular}
\caption{The ENZ nanoparticles with their characteristic ENZ wavelengths as well as their corresponding resonance wavelengths. 
}
\label{table:1}
\end{table}
This is due to the resonant excitation of dipole surface plasmons on the ENZ nanoparticle. We note, in fact, that the obtained resonance enhancement for both scattering and absorption cross-section occurs when the condition $\varepsilon_{eff} \approx -2$ (the so-called Fr\"{o}hlich condition) for spherical nanoparticles is satisfied \cite{Fan2014}. The absorption and scattering for the ENZ nanoparticle  $a_{2,1} =\{98,100\}$ nm has appreciably similar peak values.
However, for the ENZ nanoparticle  $a_{2,1} =\{38,40\}$ nm and $a_{2,1} =\{68,70\}$ nm, we obtained a relatively low scattering peak values, as compared to its corresponding absorption cross-section peak values.  It is also interesting to note that by changing the filling ratio of the ENZ nanoparticle, a passive tuning of the $\lambda_{ENZ}$ can be observed, which corroborates the shift in the nanoparticle's peak absorbance and scattering cross-sections. 
%
%
Intuitively, the redshift of the peak in the dipole resonance with increasing size could be linked to the weakening of their restoring force. Since the distance between charges on opposite sides of the ENZ nanoparticle increases with size, in fact, their corresponding interaction decrease. 

 \begin{figure}[!ht]
\includegraphics[width=1\linewidth]{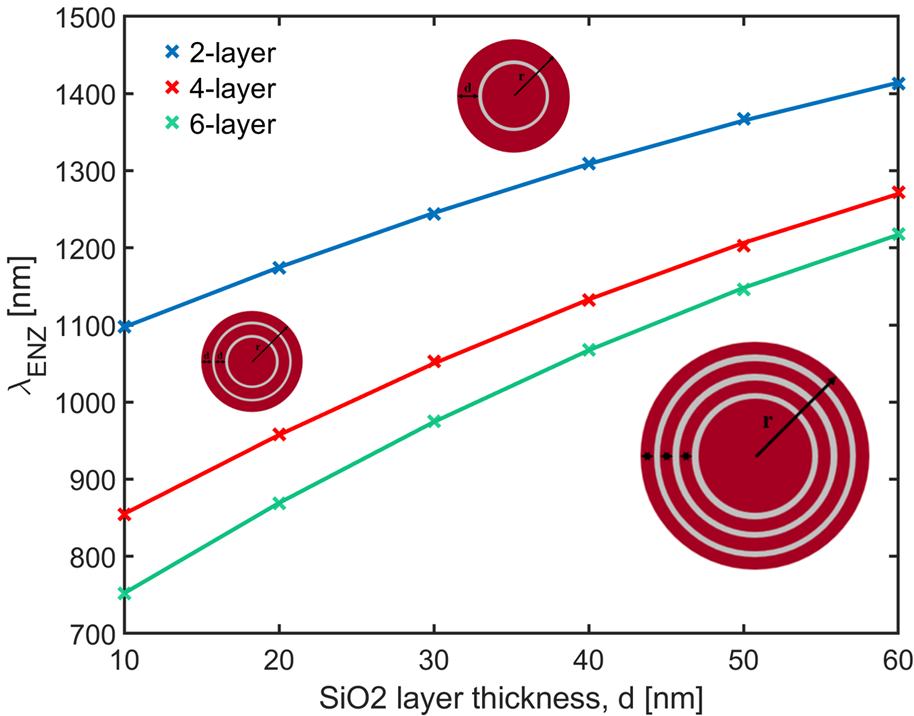}
\caption{ENZ wavelengths calculated for three different structures of a set of SiO$_{2}$ layers, with varying thicknesses (d). The SiO$_{2}$ inner core and the Ag layers in between are fixed with a radius of $a_{2} =\{98\}$ nm and a thickness of 2 nm. 
The compositions of the multilayer nanospheres are schematically presented with red (SiO$_{2}$) and gray (Ag) circular layers, next to the relevant $\lambda_{ENZ}$ plots, corresponding to 2 (blue), 4 (red), and 6 layers (green), respectively with thickness d values ranging from $\{10 - 60\}$ nm.}
\label{fig5}
\end{figure}
In addition to the absorption and scattering cross-section of the ENZ nanoparticles, we numerically calculated the ENZ wavelengths for different layered nanoparticles by parametrically varying the thickness of SiO$_{2}$ overlayed on the outer Ag shell. The SiO$_{2}$ inner core and the Ag layers in between are fixed with a radius of $a_{2} =\{98\}$ nm and a thickness of 2 nm. It is evident in Fig. \ref{fig5} that by changing the thicknesses of the embedded SiO$_{2}$ layers, we attain different ENZ wavelengths for the different multilayered nanoparticles. For all the considered nanostructures, we see a linear trend that depicts that by varying the thicknesses of the SiO$_{2}$ layers, there is a corresponding shift in $\lambda_{ENZ}$ for different layered structures. This signifies the possibility of extending the effective medium formulation of bilayered structure into multilayered nanoparticles.

\section{\label{sec: level4} Conclusion}
Our work shows that effective medium theory is potentially applicable to bilayer spherical nanoparticles to determine their unique spectral responses and ENZ properties. By changing the diameter of the ENZ nanoparticle, we identified a spectral shift in the spectral absorbance and scattering cross-sections of the ENZ nanoparticle which signifies a passive tuning of the proposed structure. The effective permittivity formulation works relatively well for bilayer structures and could be easily extended to multilayered structures as shown in subsequent discussions. The electric near-field response for both the bilayer and the effective medium structures shows similar near-field optical responses. 
Our proposed ENZ nanoparticle can be obtained by low-cost chemical synthesis techniques to be utilized in applications that grasp the advantage of ENZ properties as well nanoparticle properties.

\begin{acknowledgments}
The authors acknowledge the financial support of the Academy of Finland Flagship Programme (PREIN - decision 320165). H.C. acknowledges the financial support of the European Research Council (Starting Grant project aQUARiUM; Agreement No. 802986). I.I acknowledge Optica for the Optica Foundation Amplify Scholarship and SPIE for the SPIE Optics and Photonics Education Scholarship. 
\end{acknowledgments}

\section*{AUTHORS’ CONTRIBUTIONS}
H.C. and M.O. contributed equally to this work.


\bibliography{aipsamp}



\end{document}